\documentclass[aps,prl,twocolumn,groupedaddress]{revtex4}

\usepackage{graphicx}
\usepackage{mathrsfs}

\begin{document} 

\title{Spin torque and charge resistance of ferromagnetic semiconductor $2\pi$ and $\pi$ domain walls} 
\author{E. A. Golovatski}
\affiliation{Optical Science and Technology Center and Department of Physics and Astronomy, University of Iowa, Iowa City, IA 52242}
\author{M. E. Flatt\'e}
\affiliation{Optical Science and Technology Center and Department of Physics and Astronomy, University of Iowa, Iowa City, IA 52242}

\begin{abstract}
Charge resistance and spin torque are generated by coherent carrier transport through ferromagnetic  $2\pi$ domain walls, although they follow qualitatively different trends than for $\pi$ domain walls. The charge resistance of $2\pi$ domain walls reaches a maximum at an intermediate wall thickness, unlike  $\pi$ domain walls, whose resistance decreases monotonically with wall thickness.  The peak amplitude of the spin torque and the optimal thickness of the domain wall to maximize torque for a $2\pi$ wall are more than twice as large as found for a $\pi$ domain wall in the same material, producing a larger domain wall velocity for the $2\pi$ wall and suggesting unexpected nonlinearities in magnetoelectronic devices incorporating domain wall motion.
\end{abstract}
 





\maketitle 

Spin torque generated by spin transport through inhomogeneous magnetic systems, a direct manifestation of the conservation of the angular momentum associated with spin, underlies both unresolved fundamental questions and potential applications, including fast, localized electrical switching of magnetic moments or domains\cite{Slonczewski1996, Berger1996,Tsoi1998,Myers1999,Ozyilmaz2004}, depinning and transport of domain walls\cite{Yamanouchi2004,Waintal2004, Grollier2004,Parkin2008,Miron2009,Lepadatu2009,Tserkovnyak2008}, electrical driving of ferromagnetic resonance\cite{Kasai2006, Sankey2006, Tulapurkar2005, Fuchs2007}, and controlled generation of coherent magnons\cite{Tserkovnyak2008, Balashov2008}. Structures showing spin torque are commonly domain walls between two regions whose magnetization orientation differs by an angle $\theta$, called $\theta$-domain walls. 
Although spin torque on a $\pi$ wall has garnered much experimental and theoretical attention\cite{Yamanouchi2004, Yamanouchi2006, Yamaguchi2004, Morozovska2008, Nguyen2007,Thiaville2005, Barnes2005, Waintal2004, Xiao2006, Dugaev2006, Tatara2004, Ohno2008}, little has been done to explore spin torque in $2\pi$ walls, which are known to be stable in many metallic systems\cite{Muratov2008}, and have been seen experimentally\cite{Smith1962}.  The difference between $2\pi$ wall behavior and $\pi$ behavior might be most marked when ballistic transport across the domain wall is possible, such as for magnetic semiconductor domain walls (whose $\pi$ walls are predicted to have highly non-linear spin transport properties\cite{Vignale2002, Deutsch2004}, although spin torque was not explored).  Understanding $2\pi$ domain walls may also lead to novel spin torque devices, such have been predicted for $\pi$ walls\cite{Allwood2005, Ono2008, Parkin2008}.

Here we calculate the charge and spin transport and torque for $\pi$ and $2\pi$ domain walls 
in a ferromagnetic semiconductor.  Under the conditions of coherent transport, analytic solutions for spin-dependent transmission and reflection coefficients for the different spin channels are possible\cite{Levy1997,Vignale2002}.
Highly nonlinear voltage dependence of the spin transport and spin torque occurs for both the $\pi$ and $2\pi$ walls, but with very different wall thickness dependence. The $2\pi$ domain wall resistance vanishes in the limit of zero thickness as well as for thick walls (in which the spin adiabatically follows the local magnetization), but peaks for intermediate thicknesses; the $\pi$ domain wall resistance monotonically decreases with thickness. The spin torque on a $\pi$ wall is insensitive to domain wall width, except for very thin walls.  For $2\pi$ walls, however, a large spin torque is generated by spin transport over a range of intermediate wall widths, but  very little spin torque is generated for both very thin and very thick walls.  Even more surprising, the peak domain wall velocity is larger for a $2\pi$ wall than a $\pi$ wall, $3\pi$ wall, or $4\pi$ wall, suggesting that multiple-rotation (helical) walls may provide the fastest domain wall velocities in a ferromagnetic semiconductor material.

\begin{figure}
\includegraphics[width=\columnwidth]{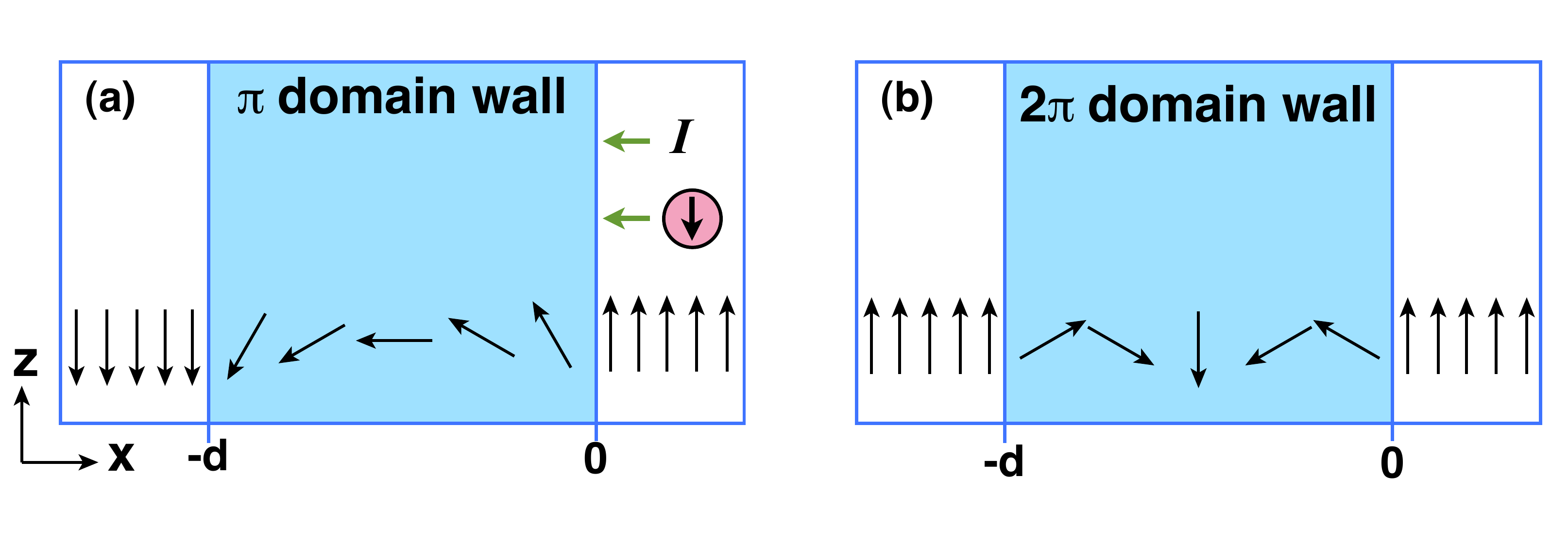}
\caption{(color online) Schematic representation of N\'eel (a) $\pi$- and (b) $2\pi$-domain walls. Charge transport is assumed to be by holes with spin antiparallel to magnetization (as in GaMnAs).}
\label{walls}
\end{figure}

Schematics of the $\pi$-domain wall and $2\pi$-domain wall are shown in Fig.~\ref{walls}.  There are two regions of ferromagnetic material, with their magnetizations oriented antiparallel for the $\pi$ wall case and parallel for the $2\pi$ wall case, separated by a N\'eel type domain wall (energetically favorable in thin films\cite{ChikazumiBook}) .  The exchange field in the domain wall is approximated to be
\begin{equation}
{\bf B} = B_0 [\sin\theta(x) {\bf \hat x} + \cos\theta(x) {\bf \hat z}],
\label{effectivefield}
\end{equation}
where $\theta$ varies smoothly with x in the form $\theta = \phi x/d$, and $\phi = \pi$ or $2\pi$ is the angle through which the magnetization rotates from $x = 0$ to $x = -d$. 

Spin transport through the domain wall begins with carriers incident on the right hand side of the domain wall, with their spins oriented antiparallel to the magnetization in that region (the case for GaMnAs, Fig.~\ref{walls}).  These carriers can be reflected or transmitted either with or without flipping their spins, and the incoming, reflected, and transmitted wavefunctions are:
\begin{eqnarray}
\psi_{in}  &=&  {\left(\begin{array}{c} e^{-i k_\uparrow x} \\ 0 \end{array}\right)},\quad
 \label{psiin}\\
\psi_{r}  &=&  {\left(\begin{array}{c} r_{nf} \  e^{i k_\uparrow x} \\ r_{sf} \  e^{i k_\downarrow x} \end{array}\right)},\quad
\psi_{t}  =  {\left(\begin{array}{c} t_{nf} \  e^{-i k_\uparrow x} \\ t_{sf} \  e^{-i k_\downarrow x} \end{array}\right)},\label{psit}
 \end{eqnarray}
where $t_{sf}$($t_{nf}$) and $r_{sf}$($r_{nf}$) are the coefficients for transmission and reflection with(without) spin flip.

We calculate the reflection and transmission coefficients by solving the Schr\"odinger equation inside the domain wall\cite{Vignale2002}
\begin{equation}
\left[{-\hbar^2 \over 2m^*}{\partial^2 \over \partial x^2} - 
{\Delta \over 2}
{\left(\begin{array}{c c} \cos\theta(x) & \sin\theta(x) \\ \sin\theta(x) & -\cos\theta(x) \end{array}\right)}\right]{\left(\begin{array}{c} \psi_\uparrow \\ \psi_\downarrow \end{array}\right)} = E {\left(\begin{array}{c} \psi_\uparrow \\ \psi_\downarrow \end{array}\right),}
\label{hamiltonian}
\end{equation}
where $\Delta$ is the energy splitting between carriers of opposite spin orientation in the ferromagnetic material.

With a position-dependent $\theta$ in the Hamiltonian, it is most convenient to transform to a rotating frame\cite{Calvo1978}.  The rotation matrix
\begin{equation}
R = e^{-{i \theta \over 2} \sigma_y} = {\left(\begin{array}{c c}\cos{\theta(x) \over 2} & \sin{\theta(x) \over 2}) \\ -\sin{\theta(x) \over 2} &\cos{\theta(x) \over 2} \end{array}\right)}
\label{rotation}
\end{equation}
defines $\psi = R\varphi$ and removes the $\theta$ dependence from the off-diagonal potential matrix:
\begin{equation}
R^{-1}{\left(\begin{array}{c c} \cos\theta(x) & \sin\theta(x) \\ \sin\theta(x) & -\cos\theta(x) \end{array}\right)}R = \sigma_z,
\label{rsigma}
\end{equation}
and yields a modified Schr\"odinger equation:
\begin{equation}
\left[{-\hbar^2 \over 2m^*}{\partial^2 \over \partial x^2} + {i\hbar^2\phi \over 2m^*d}{\sigma_y}{\partial \over \partial x} - {\Delta \over 2d^2}{\sigma_z} + {\hbar^2\phi^2 \over 8m^*d^2}\right]\varphi = {E \over d^2} \ \varphi.
\label{newSE}
\end{equation}
Eq.~(\ref{newSE}) can be solved analytically for the wavefunctions inside the domain wall.  We then set up matching conditions for the wavefunctions and their derivatives at the wall boundaries, and solve for the transmission and reflection coefficients.

After obtaining the full wavefunctions for the entire system, we define a charge current density {\bf J} and spin current density {\bf Q}\cite{Ralph2008}
\begin{eqnarray}
{\bf J} &=& {e \hbar \over 2im^{*}}[\psi^\dagger \, (\partial_x \psi) - (\partial_x \, \psi^\dagger) \, \psi]{\bf \hat x}.\label{jcurrent}\\
{\bf Q} &=& {\hbar \over 2im^{*}}[\psi^\dagger \, {\bf S} \, (\partial_x \psi) - (\partial_x \, \psi^\dagger) \, {\bf S} \, \psi]. \label{qcurrent}
\end{eqnarray}
The tensor {\bf Q} has a flow direction in real space as well as a direction in spin space.  As our transport model is one-dimensional, the real-space flow direction lies solely along the $\hat x$ direction, and we write {\bf Q} as a vector with components corresponding to the appropriate spin-space directions.
As this spin current is not a conserved quantity, we can then define the spin torque per unit area as the amount of spin current lost to the domain wall during transport\cite{Ralph2008}:
 \begin{equation}
{\bf N} = {\bf Q}_{in} + {\bf Q}_{r}- {\bf Q}_{t}.
\label{ntorque}
\end{equation}
Charge currents and spin torque can be calculated by integrating the transmission and reflection coefficients from Eq.~(\ref{newSE}) over the carrier population.

Calculations here will treat a model representative of GaMnAs, corresponding to a spin-split three-dimensional parabolic hole band with spin splitting $\Delta = 100$~meV and valence hole effective mass  $m^* = 0.45\, m_e$, where $m_e$ is the mass of the bare electron.  We assume a temperature of 110 K and a carrier density of $\sim 10^{19}$ cm$^{-3}$. For these parameters GaMnAs is effectively a 100\% spin-polarized ferromagnetic semiconductor, for which the effects we find are most visible. Although the results change quantitatively for different parameters (corresponding, {\it e.g.}, to lower spin polarization in GaMnAs), the qualitative trends we have identified are robust so long as the Fermi energy and the temperature are much less than the spin splitting ($100$~meV).

\begin{figure}
\includegraphics[width=\columnwidth]{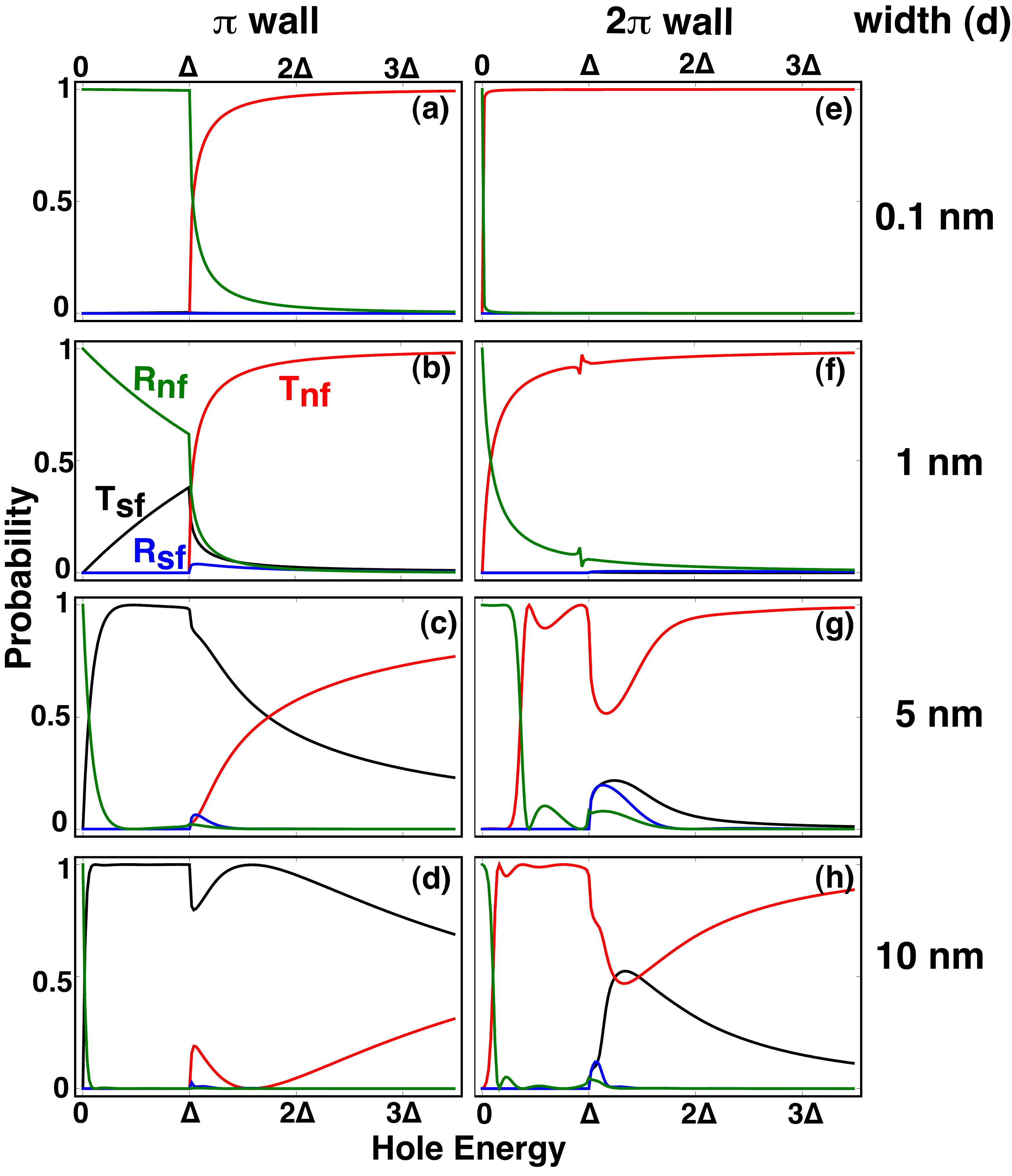}
\caption{(color online) Probabilities for transmission with spin flip  $(T_{sf})$, transmission without spin flip $(T_{nf})$, reflection with spin flip $(R_{sf})$ and reflection without spin flip $(R_{nf})$ for (a-d) $\pi$ and (e-h) $2\pi$ N\'eel walls with widths of 0.1 nm, 1 nm, 5 nm, and 10 nm  }
\label{probgrid}
\end{figure}

\begin{figure}
\includegraphics[width=\columnwidth]{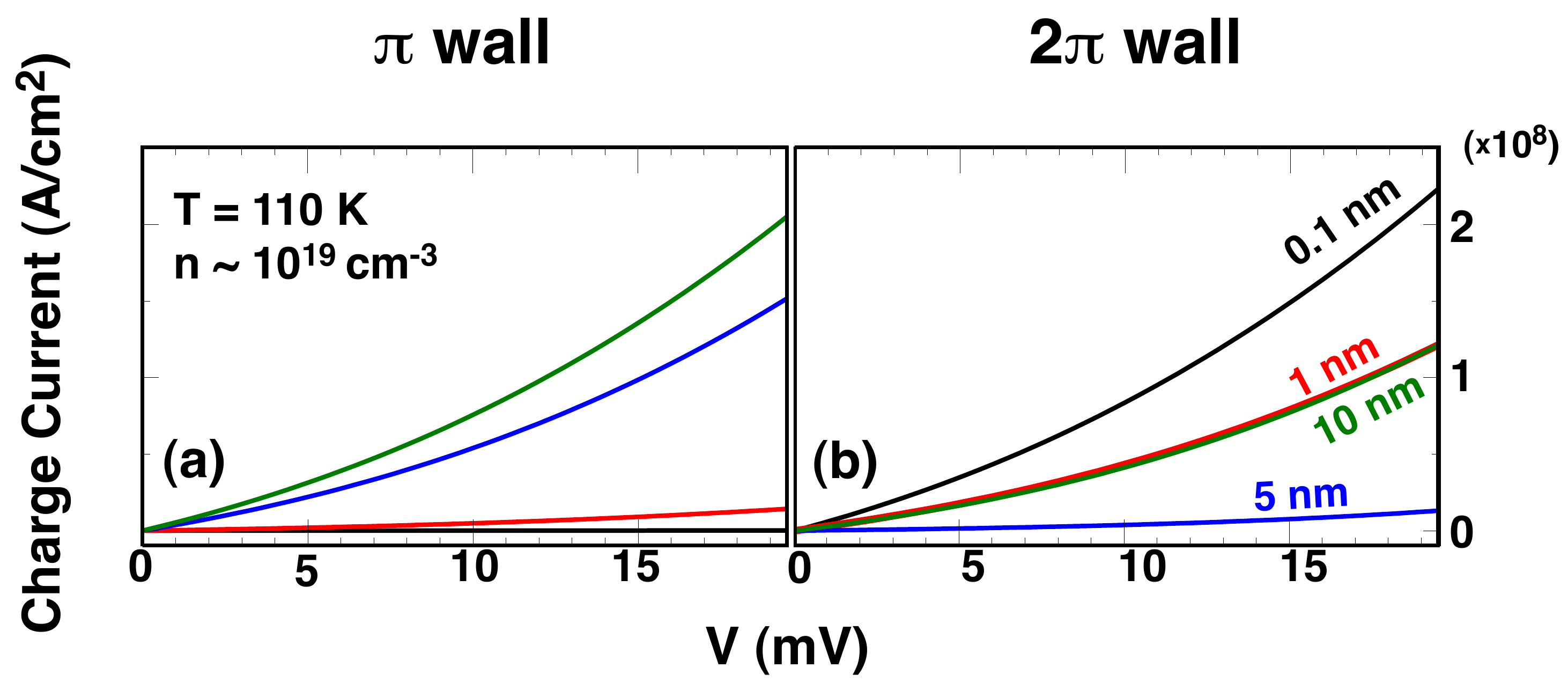}
\caption{(color online) Charge current as a function of bias voltage for $\pi$(a) and $2\pi$(b) walls.  Curves correspond to different domain wall widths. }
\label{jegrid}
\end{figure}

Fig.~\ref{probgrid} shows calculated probabilities for transmission and reflection with and without spin flip for several thicknesses of $\pi$ and $2\pi$ N\'eel walls (presumably engineered by modifying film thickness and geometric shape), and Fig.~\ref{jegrid} shows the charge current when an average over the carrier population is taken of the coefficients in Fig.~\ref{probgrid}.  The thin wall and thick wall limits for $2\pi$-domain walls differ substantially from those of $\pi$-domain walls. For thin walls the carriers effectively move through the domain wall without changing their spin orientation, which leads to carrier reflection for the $\pi$-domain wall as at low energy there are no final states on the other side with the correct spin orientation (and thus high resistance\cite{Vignale2002}). A thin $2\pi$-domain wall, however, will let the carriers through efficiently and thus have low resistance. As the thickness of a $\pi$-domain wall increases, the spin-flip transmission monotonically increases (as shown for successively wider domain walls in Fig.~\ref{probgrid}(a-d)), and the resistance drops, as shown in Fig.~\ref{jegrid}(a). As the thickness of a $2\pi$-domain wall increases initially from zero thickness, spin precession in the domain wall becomes more pronounced and carrier reflection is possible, so the resistance increases. However, in the limit of a very thick $2\pi$-domain wall the carriers will adiabatically follow the local magnetization and thus will be oriented once again parallel to the final magnetization, producing low resistance. Thus a finite thickness with maximal domain wall resistance should be expected for a $2\pi$-domain wall, as shown in Fig.~\ref{jegrid}(b).

\begin{figure}
\includegraphics[width=\columnwidth]{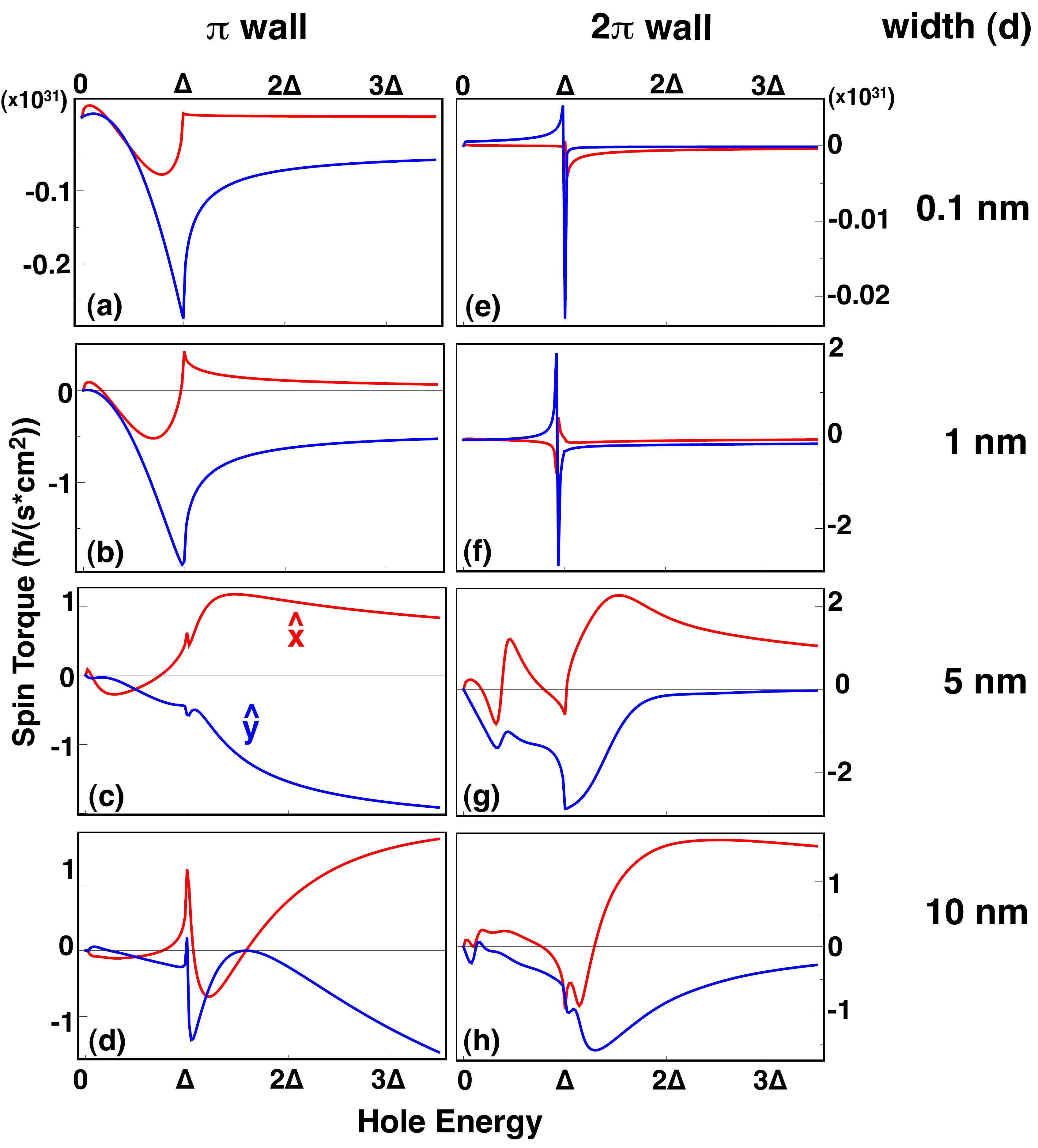}
\caption{(color online) Spin torque components as a function of hole energy for $\pi$(a-d) and $2\pi$(e-h) walls.  }
\label{energygrid}
\end{figure}

Fig.~\ref{energygrid} shows the calculated components of the spin torque from Eq.~\ref{ntorque}.  In the region of energy below the spin splitting $\Delta$, the $\pi$ wall graphs(a-d) show large $\hat x$ and $\hat y$ torque components for the 0.1 nm(a) and 1 nm(b) walls, with smaller torques for the 5 nm(c) and 10 nm(d) wall.  In the energy region below $\Delta$ for the $2\pi$ walls(e-h), we calculate almost no spin torque for the 0.1 nm(e) and 1 nm(f) walls, large $\hat x$ and $\hat y$ torque components for the 5 nm wall(g), and diminishing torques for the 10 nm wall(h).
We identify the spin torque as {\it adiabatic} (proportional to $\bf{\nabla M(r)}$, and thus parallel to $\hat x$) or {\it non-adiabatic} (proportional to $\bf{M(r)} \times \bf{\nabla M(r)}$, parallel to $\hat y$), and find both components contribute significantly to the spin torque for both $\pi$ and $2\pi$ walls. Thus the common assumption of principally adiabatic torque for $\pi$ walls in metals\cite{Zhang2004,Thiaville2005,Xiao2006} (except for very narrow walls) breaks down for coherent transport, such as it does for unusual shape anisotropy or strong spin-orbit interaction\cite{Boulle2008,Garate2009}.

\begin{figure}
\includegraphics[width=\columnwidth]{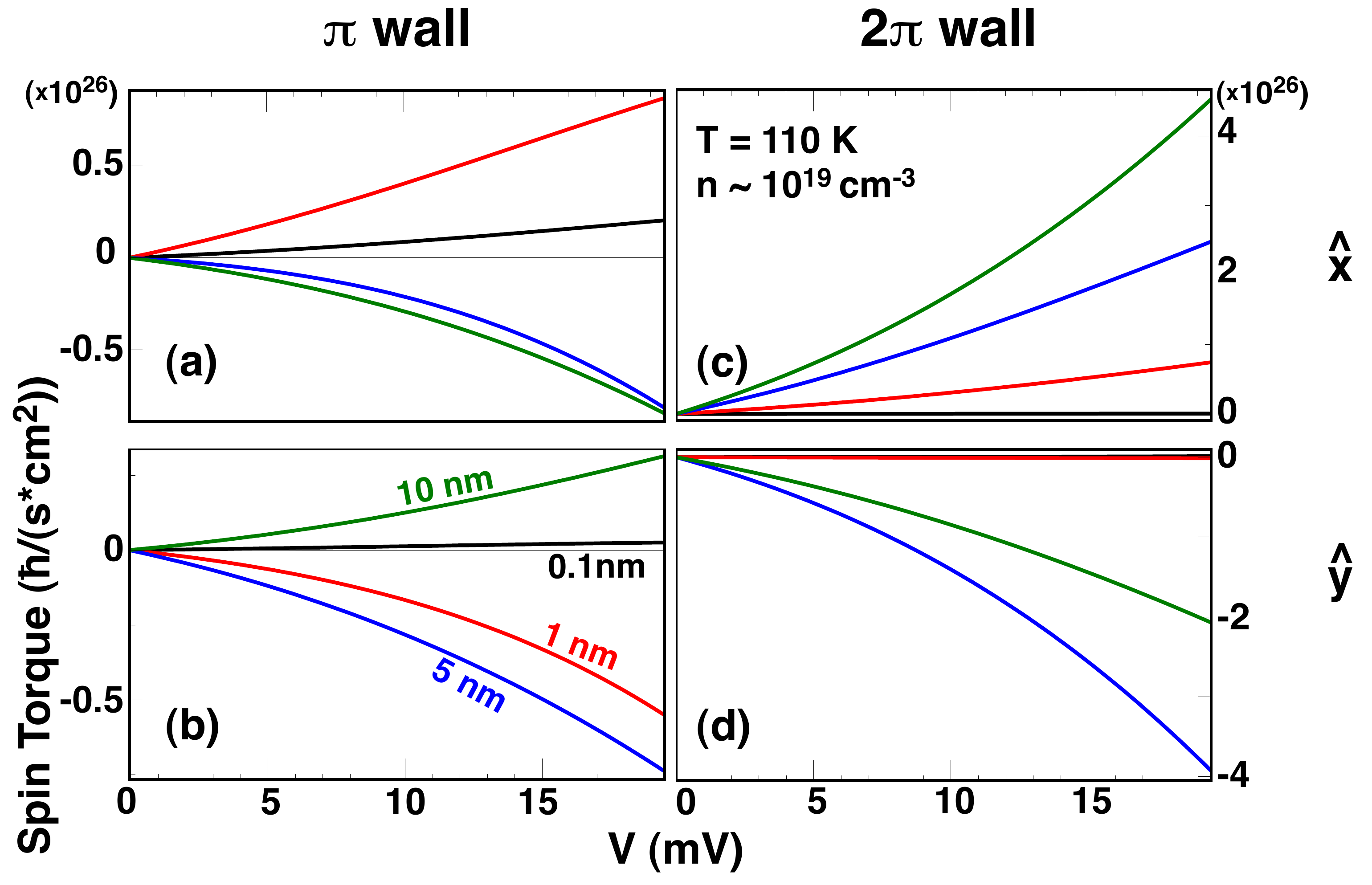}
\caption{(color online) Spin torque components as a function of bias voltage for $\pi$(a-b) and $2\pi$(c-d) walls.}
\label{ngrid}
\end{figure}

Fig.~\ref{ngrid} shows the components of the total spin torque for each type of domain wall.
 For the $\pi$ wall both torque components change sign with voltage, as higher-energy regions of Fig.~\ref{energygrid} with opposite sign torque are accessed. For the $2\pi$ wall we calculate almost no torque for the 0.1 nm and 1 nm walls, the largest of the four torques is at 5 nm (the width where the charge current has its smallest value) and a smaller torque at 10 nm.

\begin{figure}
\includegraphics[width=\columnwidth]{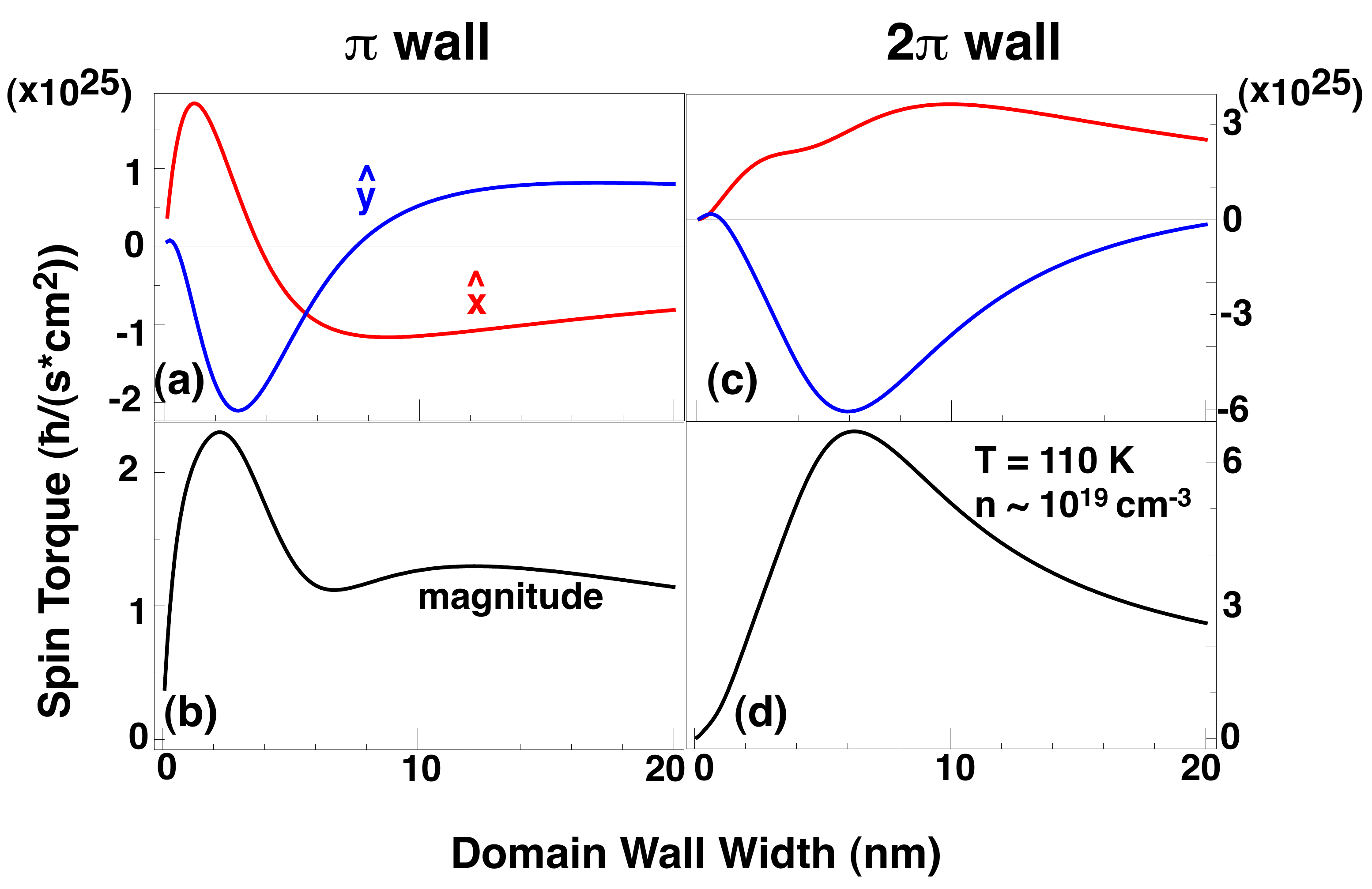}
\caption{(color online) Spin torque components and magnitudes as a function of domain wall width for a $\pi$ and $2\pi$ wall at V = 5 mV. }
\label{dgrid}
\end{figure}

Fig.~\ref{dgrid} examines this width dependence more closely, showing calculations for the individual torque components as well as the overall magnitude of the spin torque as a function of the domain wall width for a fixed applied voltage across the walls, $5$~mV.
For the $2\pi$ wall(c,d) the 
$\hat x$ component and $\hat y$ components exhibit non-trivial dependence on the width of the domain wall, peaking at widths $\sim9$ nm and $\sim6$ nm respectively.  In (d) we see that the overall magnitude of the spin torque follows the curve of the larger non-adiabatic $\hat y$ component, thus also peaking near 6 nm.

This treatment of coherent transport across domain walls has shown that the behavior of spin transport through domain walls is intrinsically nonlinear in voltage and magnetization rotation angle, for spin transport and torque both depend nonlinearly on the applied voltage, and the properties of a wall with twice the magnetization rotation ($2\pi$ wall) are not related in any clear fashion to the properties of the $\pi$ wall. The domain wall resistance to charge current follows different qualitative trends for the $2\pi$ domain wall than the $\pi$ domain wall, leading to a maximum resistance at an intermediate wall thickness, as opposed to maximum resistance at zero thickness. An optimal-thickness $2\pi$ N\'eel wall experiences more than twice as much spin torque as an optimal thickness $\pi$ domain wall for the same applied voltage, producing a domain wall velocity 50\% higher for the $2\pi$ wall.    This indicates that there an optimal width for achieving a maximum amount of spin torque, which should assist in understanding the time-dependent properties of domain walls in the presence of current, including potentially finding the fastest racers around a magnetic racetrack memory\cite{Parkin2008}.

This work was supported by an ARO MURI.


\begin{thebibliography}{40}
\expandafter\ifx\csname natexlab\endcsname\relax\def\natexlab#1{#1}\fi
\expandafter\ifx\csname bibnamefont\endcsname\relax
  \def\bibnamefont#1{#1}\fi
\expandafter\ifx\csname bibfnamefont\endcsname\relax
  \def\bibfnamefont#1{#1}\fi
\expandafter\ifx\csname citenamefont\endcsname\relax
  \def\citenamefont#1{#1}\fi
\expandafter\ifx\csname url\endcsname\relax
  \def\url#1{\texttt{#1}}\fi
\expandafter\ifx\csname urlprefix\endcsname\relax\def\urlprefix{URL }\fi
\providecommand{\bibinfo}[2]{#2}
\providecommand{\eprint}[2][]{\url{#2}}

\bibitem[{\citenamefont{Slonczewski}(1996)}]{Slonczewski1996}
\bibinfo{author}{\bibfnamefont{J.}~\bibnamefont{Slonczewski}},
  \bibinfo{journal}{Journal of Magnetism and Magnetic Materials}
  \textbf{\bibinfo{volume}{159}}, \bibinfo{pages}{L1} (\bibinfo{year}{1996}).

\bibitem[{\citenamefont{Berger}(1996)}]{Berger1996}
\bibinfo{author}{\bibfnamefont{L.}~\bibnamefont{Berger}},
  \bibinfo{journal}{Phys. Rev. B} \textbf{\bibinfo{volume}{54}},
  \bibinfo{pages}{9353} (\bibinfo{year}{1996}).

\bibitem[{\citenamefont{Tsoi et~al.}(1998)\citenamefont{Tsoi, Jansen, Bass,
  Chiang, Seck, Tsoi, and Wyder}}]{Tsoi1998}
\bibinfo{author}{\bibfnamefont{M.}~\bibnamefont{Tsoi}},
  \bibinfo{author}{\bibfnamefont{A.~G.~M.} \bibnamefont{Jansen}},
  \bibinfo{author}{\bibfnamefont{J.}~\bibnamefont{Bass}},
  \bibinfo{author}{\bibfnamefont{W.-C.} \bibnamefont{Chiang}},
  \bibinfo{author}{\bibfnamefont{M.}~\bibnamefont{Seck}},
  \bibinfo{author}{\bibfnamefont{V.}~\bibnamefont{Tsoi}}, \bibnamefont{and}
  \bibinfo{author}{\bibfnamefont{P.}~\bibnamefont{Wyder}},
  \bibinfo{journal}{Phys. Rev. Lett.} \textbf{\bibinfo{volume}{80}},
  \bibinfo{pages}{4281} (\bibinfo{year}{1998}).

\bibitem[{\citenamefont{Myers et~al.}(1999)\citenamefont{Myers, Ralph, Katine,
  Louie, and Buhrman}}]{Myers1999}
\bibinfo{author}{\bibfnamefont{E.~B.} \bibnamefont{Myers}},
  \bibinfo{author}{\bibfnamefont{D.~C.} \bibnamefont{Ralph}},
  \bibinfo{author}{\bibfnamefont{J.~A.} \bibnamefont{Katine}},
  \bibinfo{author}{\bibfnamefont{R.~N.} \bibnamefont{Louie}}, \bibnamefont{and}
  \bibinfo{author}{\bibfnamefont{R.~A.} \bibnamefont{Buhrman}},
  \bibinfo{journal}{Science} \textbf{\bibinfo{volume}{285}},
  \bibinfo{pages}{867} (\bibinfo{year}{1999}).

\bibitem[{\citenamefont{\"Ozyilmaz et~al.}(2004)\citenamefont{\"Ozyilmaz, Kent,
  Sun, Rooks, and Koch}}]{Ozyilmaz2004}
\bibinfo{author}{\bibfnamefont{B.}~\bibnamefont{\"Ozyilmaz}},
  \bibinfo{author}{\bibfnamefont{A.~D.} \bibnamefont{Kent}},
  \bibinfo{author}{\bibfnamefont{J.~Z.} \bibnamefont{Sun}},
  \bibinfo{author}{\bibfnamefont{M.~J.} \bibnamefont{Rooks}}, \bibnamefont{and}
  \bibinfo{author}{\bibfnamefont{R.~H.} \bibnamefont{Koch}},
  \bibinfo{journal}{Phys. Rev. Lett.} \textbf{\bibinfo{volume}{93}},
  \bibinfo{pages}{176604} (\bibinfo{year}{2004}).

\bibitem[{\citenamefont{Yamanouchi et~al.}(2004)\citenamefont{Yamanouchi,
  Chiba, Matsukura, and Ohno}}]{Yamanouchi2004}
\bibinfo{author}{\bibfnamefont{M.}~\bibnamefont{Yamanouchi}},
  \bibinfo{author}{\bibfnamefont{D.}~\bibnamefont{Chiba}},
  \bibinfo{author}{\bibfnamefont{F.}~\bibnamefont{Matsukura}},
  \bibnamefont{and} \bibinfo{author}{\bibfnamefont{H.}~\bibnamefont{Ohno}},
  \bibinfo{journal}{Nature} \textbf{\bibinfo{volume}{428}},
  \bibinfo{pages}{539} (\bibinfo{year}{2004}).

\bibitem[{\citenamefont{Waintal and Viret}(2004)}]{Waintal2004}
\bibinfo{author}{\bibfnamefont{X.}~\bibnamefont{Waintal}} \bibnamefont{and}
  \bibinfo{author}{\bibfnamefont{M.}~\bibnamefont{Viret}},
  \bibinfo{journal}{Europhysics Lett.} \textbf{\bibinfo{volume}{65}},
  \bibinfo{pages}{427} (\bibinfo{year}{2004}).

\bibitem[{\citenamefont{Grollier et~al.}(2004)\citenamefont{Grollier, Boulenc,
  Cros, Hamzi\'c, Vaur\`es, and Fert}}]{Grollier2004}
\bibinfo{author}{\bibfnamefont{J.}~\bibnamefont{Grollier}},
  \bibinfo{author}{\bibfnamefont{P.}~\bibnamefont{Boulenc}},
  \bibinfo{author}{\bibfnamefont{V.}~\bibnamefont{Cros}},
  \bibinfo{author}{\bibfnamefont{A.}~\bibnamefont{Hamzi\'c}},
  \bibinfo{author}{\bibfnamefont{A.}~\bibnamefont{Vaur\`es}}, \bibnamefont{and}
  \bibinfo{author}{\bibfnamefont{A.}~\bibnamefont{Fert}}, \bibinfo{journal}{J.
  Appl. Phys.} \textbf{\bibinfo{volume}{95}}, \bibinfo{pages}{6777}
  (\bibinfo{year}{2004}).

\bibitem[{\citenamefont{Parkin et~al.}(2008)\citenamefont{Parkin, Hayashi, and
  Thomas}}]{Parkin2008}
\bibinfo{author}{\bibfnamefont{S.~S.~P.} \bibnamefont{Parkin}},
  \bibinfo{author}{\bibfnamefont{M.}~\bibnamefont{Hayashi}}, \bibnamefont{and}
  \bibinfo{author}{\bibfnamefont{L.}~\bibnamefont{Thomas}},
  \bibinfo{journal}{Science} \textbf{\bibinfo{volume}{320}},
  \bibinfo{pages}{190} (\bibinfo{year}{2008}).

\bibitem[{\citenamefont{Miron et~al.}(2009)\citenamefont{Miron, Zermatten,
  Gaudin, Auffret, Rodmacq, and Schuhl}}]{Miron2009}
\bibinfo{author}{\bibfnamefont{I.~M.} \bibnamefont{Miron}},
  \bibinfo{author}{\bibfnamefont{P.-J.} \bibnamefont{Zermatten}},
  \bibinfo{author}{\bibfnamefont{G.}~\bibnamefont{Gaudin}},
  \bibinfo{author}{\bibfnamefont{S.}~\bibnamefont{Auffret}},
  \bibinfo{author}{\bibfnamefont{B.}~\bibnamefont{Rodmacq}}, \bibnamefont{and}
  \bibinfo{author}{\bibfnamefont{A.}~\bibnamefont{Schuhl}},
  \bibinfo{journal}{Phys. Rev. Lett.} \textbf{\bibinfo{volume}{102}},
  \bibinfo{pages}{137202} (\bibinfo{year}{2009}).

\bibitem[{\citenamefont{Lepadatu et~al.}(2009)\citenamefont{Lepadatu,
  Vanhaverbeke, Atkinson, Allenspach, and Marrows}}]{Lepadatu2009}
\bibinfo{author}{\bibfnamefont{S.}~\bibnamefont{Lepadatu}},
  \bibinfo{author}{\bibfnamefont{A.}~\bibnamefont{Vanhaverbeke}},
  \bibinfo{author}{\bibfnamefont{D.}~\bibnamefont{Atkinson}},
  \bibinfo{author}{\bibfnamefont{R.}~\bibnamefont{Allenspach}},
  \bibnamefont{and} \bibinfo{author}{\bibfnamefont{C.~H.}
  \bibnamefont{Marrows}}, \bibinfo{journal}{Phys. Rev. Lett.}
  \textbf{\bibinfo{volume}{102}}, \bibinfo{pages}{127203}
  (\bibinfo{year}{2009}).

\bibitem[{\citenamefont{Tserkovnyak et~al.}(2008)\citenamefont{Tserkovnyak,
  Brataas, and Bauer}}]{Tserkovnyak2008}
\bibinfo{author}{\bibfnamefont{T.}~\bibnamefont{Tserkovnyak}},
  \bibinfo{author}{\bibfnamefont{A.}~\bibnamefont{Brataas}}, \bibnamefont{and}
  \bibinfo{author}{\bibfnamefont{G.}~\bibnamefont{Bauer}},
  \bibinfo{journal}{Journal of Magnetism and Magnetic Materials}
  \textbf{\bibinfo{volume}{320}}, \bibinfo{pages}{1282} (\bibinfo{year}{2008}).

\bibitem[{\citenamefont{Kasai et~al.}(2006)\citenamefont{Kasai, Nakatani,
  Kobayashi, Kohno, and Ono}}]{Kasai2006}
\bibinfo{author}{\bibfnamefont{S.}~\bibnamefont{Kasai}},
  \bibinfo{author}{\bibfnamefont{Y.}~\bibnamefont{Nakatani}},
  \bibinfo{author}{\bibfnamefont{K.}~\bibnamefont{Kobayashi}},
  \bibinfo{author}{\bibfnamefont{H.}~\bibnamefont{Kohno}}, \bibnamefont{and}
  \bibinfo{author}{\bibfnamefont{T.}~\bibnamefont{Ono}},
  \bibinfo{journal}{Phys. Rev. Lett.} \textbf{\bibinfo{volume}{97}},
  \bibinfo{pages}{107204} (\bibinfo{year}{2006}).

\bibitem[{\citenamefont{Sankey et~al.}(2006)\citenamefont{Sankey, Braganca,
  Garcia, Krivorotov, Buhrman, and Ralph}}]{Sankey2006}
\bibinfo{author}{\bibfnamefont{J.~C.} \bibnamefont{Sankey}},
  \bibinfo{author}{\bibfnamefont{P.~M.} \bibnamefont{Braganca}},
  \bibinfo{author}{\bibfnamefont{A.~G.~F.} \bibnamefont{Garcia}},
  \bibinfo{author}{\bibfnamefont{I.~N.} \bibnamefont{Krivorotov}},
  \bibinfo{author}{\bibfnamefont{R.~A.} \bibnamefont{Buhrman}},
  \bibnamefont{and} \bibinfo{author}{\bibfnamefont{D.~C.} \bibnamefont{Ralph}},
  \bibinfo{journal}{Phys. Rev. Lett.} \textbf{\bibinfo{volume}{96}},
  \bibinfo{pages}{227601} (\bibinfo{year}{2006}).

\bibitem[{\citenamefont{Tulapurkar et~al.}(2005)\citenamefont{Tulapurkar,
  Suzuki, Fukushima, Kubota, Maehara, Tsunekawa, Djayaprawira, Watanabe, and
  Yuasa}}]{Tulapurkar2005}
\bibinfo{author}{\bibfnamefont{A.}~\bibnamefont{Tulapurkar}},
  \bibinfo{author}{\bibfnamefont{Y.}~\bibnamefont{Suzuki}},
  \bibinfo{author}{\bibfnamefont{A.}~\bibnamefont{Fukushima}},
  \bibinfo{author}{\bibfnamefont{H.}~\bibnamefont{Kubota}},
  \bibinfo{author}{\bibfnamefont{H.}~\bibnamefont{Maehara}},
  \bibinfo{author}{\bibfnamefont{K.}~\bibnamefont{Tsunekawa}},
  \bibinfo{author}{\bibfnamefont{D.}~\bibnamefont{Djayaprawira}},
  \bibinfo{author}{\bibfnamefont{N.}~\bibnamefont{Watanabe}}, \bibnamefont{and}
  \bibinfo{author}{\bibfnamefont{S.}~\bibnamefont{Yuasa}},
  \bibinfo{journal}{Nature} \textbf{\bibinfo{volume}{438}},
  \bibinfo{pages}{339} (\bibinfo{year}{2005}).

\bibitem[{\citenamefont{Fuchs et~al.}(2007)\citenamefont{Fuchs, Sankey,
  Pribiag, Qian, Braganca, Garcia, Ryan, Li, Ralph, and Buhrman}}]{Fuchs2007}
\bibinfo{author}{\bibfnamefont{G.~D.} \bibnamefont{Fuchs}},
  \bibinfo{author}{\bibfnamefont{J.~C.} \bibnamefont{Sankey}},
  \bibinfo{author}{\bibfnamefont{V.~S.} \bibnamefont{Pribiag}},
  \bibinfo{author}{\bibfnamefont{L.}~\bibnamefont{Qian}},
  \bibinfo{author}{\bibfnamefont{P.~M.} \bibnamefont{Braganca}},
  \bibinfo{author}{\bibfnamefont{A.~G.~F.} \bibnamefont{Garcia}},
  \bibinfo{author}{\bibfnamefont{E.~M.} \bibnamefont{Ryan}},
  \bibinfo{author}{\bibfnamefont{Z.-P.} \bibnamefont{Li}},
  \bibinfo{author}{\bibfnamefont{D.~C.} \bibnamefont{Ralph}}, \bibnamefont{and}
  \bibinfo{author}{\bibfnamefont{R.~A.} \bibnamefont{Buhrman}},
  \bibinfo{journal}{Applied Physics Letters} \textbf{\bibinfo{volume}{91}},
  \bibinfo{pages}{062507} (\bibinfo{year}{2007}).

\bibitem[{\citenamefont{Balashov et~al.}(2008)\citenamefont{Balashov, Tak\'acs,
  D\"ane, Ernst, Bruno, and Wulfhekel}}]{Balashov2008}
\bibinfo{author}{\bibfnamefont{T.}~\bibnamefont{Balashov}},
  \bibinfo{author}{\bibfnamefont{A.~F.} \bibnamefont{Tak\'acs}},
  \bibinfo{author}{\bibfnamefont{M.}~\bibnamefont{D\"ane}},
  \bibinfo{author}{\bibfnamefont{A.}~\bibnamefont{Ernst}},
  \bibinfo{author}{\bibfnamefont{P.}~\bibnamefont{Bruno}}, \bibnamefont{and}
  \bibinfo{author}{\bibfnamefont{W.}~\bibnamefont{Wulfhekel}},
  \bibinfo{journal}{Phys. Rev. B} \textbf{\bibinfo{volume}{78}},
  \bibinfo{pages}{174404} (\bibinfo{year}{2008}).

\bibitem[{\citenamefont{Yamanouchi et~al.}(2006)\citenamefont{Yamanouchi,
  Chiba, Matsukura, Dietl, and Ohno}}]{Yamanouchi2006}
\bibinfo{author}{\bibfnamefont{M.}~\bibnamefont{Yamanouchi}},
  \bibinfo{author}{\bibfnamefont{D.}~\bibnamefont{Chiba}},
  \bibinfo{author}{\bibfnamefont{F.}~\bibnamefont{Matsukura}},
  \bibinfo{author}{\bibfnamefont{T.}~\bibnamefont{Dietl}}, \bibnamefont{and}
  \bibinfo{author}{\bibfnamefont{H.}~\bibnamefont{Ohno}},
  \bibinfo{journal}{Phys. Rev. Lett.} \textbf{\bibinfo{volume}{96}},
  \bibinfo{pages}{096601} (\bibinfo{year}{2006}).

\bibitem[{\citenamefont{Yamaguchi et~al.}(2004)\citenamefont{Yamaguchi, Ono,
  Nasu, Miyake, Mibu, and Shinjo}}]{Yamaguchi2004}
\bibinfo{author}{\bibfnamefont{A.}~\bibnamefont{Yamaguchi}},
  \bibinfo{author}{\bibfnamefont{T.}~\bibnamefont{Ono}},
  \bibinfo{author}{\bibfnamefont{S.}~\bibnamefont{Nasu}},
  \bibinfo{author}{\bibfnamefont{K.}~\bibnamefont{Miyake}},
  \bibinfo{author}{\bibfnamefont{K.}~\bibnamefont{Mibu}}, \bibnamefont{and}
  \bibinfo{author}{\bibfnamefont{T.}~\bibnamefont{Shinjo}},
  \bibinfo{journal}{Phys. Rev. Lett.} \textbf{\bibinfo{volume}{92}},
  \bibinfo{pages}{077205} (\bibinfo{year}{2004}).

\bibitem[{\citenamefont{Morozovska et~al.}(2008)\citenamefont{Morozovska,
  Kalinin, Eliseev, Gopalan, and Svechnikov}}]{Morozovska2008}
\bibinfo{author}{\bibfnamefont{A.~N.} \bibnamefont{Morozovska}},
  \bibinfo{author}{\bibfnamefont{S.~V.} \bibnamefont{Kalinin}},
  \bibinfo{author}{\bibfnamefont{E.~A.} \bibnamefont{Eliseev}},
  \bibinfo{author}{\bibfnamefont{V.}~\bibnamefont{Gopalan}}, \bibnamefont{and}
  \bibinfo{author}{\bibfnamefont{S.~V.} \bibnamefont{Svechnikov}},
  \bibinfo{journal}{Phys. Rev. B} \textbf{\bibinfo{volume}{78}},
  \bibinfo{pages}{125407} (\bibinfo{year}{2008}).

\bibitem[{\citenamefont{Nguyen et~al.}(2007)\citenamefont{Nguyen, Skadsem, and
  Brataas}}]{Nguyen2007}
\bibinfo{author}{\bibfnamefont{A.~K.} \bibnamefont{Nguyen}},
  \bibinfo{author}{\bibfnamefont{H.~J.} \bibnamefont{Skadsem}},
  \bibnamefont{and} \bibinfo{author}{\bibfnamefont{A.}~\bibnamefont{Brataas}},
  \bibinfo{journal}{Phys. Rev. Lett.} \textbf{\bibinfo{volume}{98}},
  \bibinfo{pages}{146602} (\bibinfo{year}{2007}).

\bibitem[{\citenamefont{Thiaville et~al.}(2005)\citenamefont{Thiaville,
  Nakatani, Miltat, and Suzuki}}]{Thiaville2005}
\bibinfo{author}{\bibfnamefont{A.}~\bibnamefont{Thiaville}},
  \bibinfo{author}{\bibfnamefont{Y.}~\bibnamefont{Nakatani}},
  \bibinfo{author}{\bibfnamefont{J.}~\bibnamefont{Miltat}}, \bibnamefont{and}
  \bibinfo{author}{\bibfnamefont{Y.}~\bibnamefont{Suzuki}},
  \bibinfo{journal}{Europhys. Lett.} p. \bibinfo{pages}{990}
  (\bibinfo{year}{2005}).

\bibitem[{\citenamefont{Barnes and Maekawa}(2005)}]{Barnes2005}
\bibinfo{author}{\bibfnamefont{S.~E.} \bibnamefont{Barnes}} \bibnamefont{and}
  \bibinfo{author}{\bibfnamefont{S.}~\bibnamefont{Maekawa}},
  \bibinfo{journal}{Phys. Rev. Lett.} \textbf{\bibinfo{volume}{95}},
  \bibinfo{pages}{107204} (\bibinfo{year}{2005}).

\bibitem[{\citenamefont{Xiao et~al.}(2006)\citenamefont{Xiao, Zangwill, and
  Stiles}}]{Xiao2006}
\bibinfo{author}{\bibfnamefont{J.}~\bibnamefont{Xiao}},
  \bibinfo{author}{\bibfnamefont{A.}~\bibnamefont{Zangwill}}, \bibnamefont{and}
  \bibinfo{author}{\bibfnamefont{M.~D.} \bibnamefont{Stiles}},
  \bibinfo{journal}{Phys. Rev. B} \textbf{\bibinfo{volume}{73}},
  \bibinfo{pages}{054428} (\bibinfo{year}{2006}).

\bibitem[{\citenamefont{Dugaev et~al.}(2006)\citenamefont{Dugaev, Vieira,
  Sacramento, Barn\'as, Ara\'ujo, and Berakdar}}]{Dugaev2006}
\bibinfo{author}{\bibfnamefont{V.~K.} \bibnamefont{Dugaev}},
  \bibinfo{author}{\bibfnamefont{V.~R.} \bibnamefont{Vieira}},
  \bibinfo{author}{\bibfnamefont{P.~D.} \bibnamefont{Sacramento}},
  \bibinfo{author}{\bibfnamefont{J.}~\bibnamefont{Barn\'as}},
  \bibinfo{author}{\bibfnamefont{M.~A.~N.} \bibnamefont{Ara\'ujo}},
  \bibnamefont{and} \bibinfo{author}{\bibfnamefont{J.}~\bibnamefont{Berakdar}},
  \bibinfo{journal}{Phys. Rev. B} \textbf{\bibinfo{volume}{74}},
  \bibinfo{pages}{054403} (\bibinfo{year}{2006}).

\bibitem[{\citenamefont{Tatara and Kohno}(2004)}]{Tatara2004}
\bibinfo{author}{\bibfnamefont{G.}~\bibnamefont{Tatara}} \bibnamefont{and}
  \bibinfo{author}{\bibfnamefont{H.}~\bibnamefont{Kohno}},
  \bibinfo{journal}{Phys. Rev. Lett.} \textbf{\bibinfo{volume}{92}},
  \bibinfo{pages}{086601} (\bibinfo{year}{2004}).

\bibitem[{\citenamefont{Ohno and Dietl}(2008)}]{Ohno2008}
\bibinfo{author}{\bibfnamefont{H.}~\bibnamefont{Ohno}} \bibnamefont{and}
  \bibinfo{author}{\bibfnamefont{T.}~\bibnamefont{Dietl}},
  \bibinfo{journal}{Journal of Magnetism and Magnetic Materials}
  \textbf{\bibinfo{volume}{320}}, \bibinfo{pages}{1293} (\bibinfo{year}{2008}).

\bibitem[{\citenamefont{Muratov and Osipov}(2008)}]{Muratov2008}
\bibinfo{author}{\bibfnamefont{C.}~\bibnamefont{Muratov}} \bibnamefont{and}
  \bibinfo{author}{\bibfnamefont{V.}~\bibnamefont{Osipov}},
  \bibinfo{journal}{J. Appl. Phys.} \textbf{\bibinfo{volume}{104}},
  \bibinfo{pages}{053908} (\bibinfo{year}{2008}).

\bibitem[{\citenamefont{Smith and Harte}(1962)}]{Smith1962}
\bibinfo{author}{\bibfnamefont{D.}~\bibnamefont{Smith}} \bibnamefont{and}
  \bibinfo{author}{\bibfnamefont{K.}~\bibnamefont{Harte}}, \bibinfo{journal}{J.
  Appl. Phys.} \textbf{\bibinfo{volume}{33}}, \bibinfo{pages}{1399}
  (\bibinfo{year}{1962}).

\bibitem[{\citenamefont{Vignale and Flatt\'e}(2002)}]{Vignale2002}
\bibinfo{author}{\bibfnamefont{G.}~\bibnamefont{Vignale}} \bibnamefont{and}
  \bibinfo{author}{\bibfnamefont{M.~E.} \bibnamefont{Flatt\'e}},
  \bibinfo{journal}{Phys. Rev. Lett.} \textbf{\bibinfo{volume}{89}},
  \bibinfo{pages}{098302} (\bibinfo{year}{2002}).

\bibitem[{\citenamefont{Deutsch et~al.}(2004)\citenamefont{Deutsch, Vignale,
  and Flatt\'e}}]{Deutsch2004}
\bibinfo{author}{\bibfnamefont{M.}~\bibnamefont{Deutsch}},
  \bibinfo{author}{\bibfnamefont{G.}~\bibnamefont{Vignale}}, \bibnamefont{and}
  \bibinfo{author}{\bibfnamefont{M.}~\bibnamefont{Flatt\'e}},
  \bibinfo{journal}{J. Appl. Phys.} \textbf{\bibinfo{volume}{96}}
  (\bibinfo{year}{2004}).

\bibitem[{\citenamefont{Allwood et~al.}(2005)\citenamefont{Allwood, Xiong,
  Faulkner, Atkinson, Petit, and Cowburn}}]{Allwood2005}
\bibinfo{author}{\bibfnamefont{D.}~\bibnamefont{Allwood}},
  \bibinfo{author}{\bibfnamefont{G.}~\bibnamefont{Xiong}},
  \bibinfo{author}{\bibfnamefont{C.}~\bibnamefont{Faulkner}},
  \bibinfo{author}{\bibfnamefont{D.}~\bibnamefont{Atkinson}},
  \bibinfo{author}{\bibfnamefont{D.}~\bibnamefont{Petit}}, \bibnamefont{and}
  \bibinfo{author}{\bibfnamefont{R.}~\bibnamefont{Cowburn}},
  \bibinfo{journal}{Science} \textbf{\bibinfo{volume}{309}},
  \bibinfo{pages}{1688} (\bibinfo{year}{2005}).

\bibitem[{\citenamefont{Ono and Nakatani}(2008)}]{Ono2008}
\bibinfo{author}{\bibfnamefont{T.}~\bibnamefont{Ono}} \bibnamefont{and}
  \bibinfo{author}{\bibfnamefont{Y.}~\bibnamefont{Nakatani}},
  \bibinfo{journal}{Applied Physics Express} \textbf{\bibinfo{volume}{1}},
  \bibinfo{pages}{061301} (\bibinfo{year}{2008}).

\bibitem[{\citenamefont{Levy and Zhang}(1997)}]{Levy1997}
\bibinfo{author}{\bibfnamefont{P.~M.} \bibnamefont{Levy}} \bibnamefont{and}
  \bibinfo{author}{\bibfnamefont{S.}~\bibnamefont{Zhang}},
  \bibinfo{journal}{Phys. Rev. Lett.} \textbf{\bibinfo{volume}{79}},
  \bibinfo{pages}{5110} (\bibinfo{year}{1997}).

\bibitem[{\citenamefont{Chikaaumi}(1997)}]{ChikazumiBook}
\bibinfo{author}{\bibfnamefont{S.}~\bibnamefont{Chikaaumi}},
  \emph{\bibinfo{title}{Physics of Ferromagnetism Second Edition}},
  vol.~\bibinfo{volume}{94} of \emph{\bibinfo{series}{International Series of
  Monographs on Physics}} (\bibinfo{publisher}{Oxford}, \bibinfo{year}{1997}).

\bibitem[{\citenamefont{Calvo}(1978)}]{Calvo1978}
\bibinfo{author}{\bibfnamefont{M.}~\bibnamefont{Calvo}},
  \bibinfo{journal}{Phys. Rev. B} \textbf{\bibinfo{volume}{18}},
  \bibinfo{pages}{5073} (\bibinfo{year}{1978}).

\bibitem[{\citenamefont{Ralph and Stiles}(2008)}]{Ralph2008}
\bibinfo{author}{\bibfnamefont{D.}~\bibnamefont{Ralph}} \bibnamefont{and}
  \bibinfo{author}{\bibfnamefont{M.}~\bibnamefont{Stiles}},
  \bibinfo{journal}{Journal of Magnetism and Magnetic Materials}
  \textbf{\bibinfo{volume}{320}}, \bibinfo{pages}{1190} (\bibinfo{year}{2008}).

\bibitem[{\citenamefont{Zhang and Li}(2004)}]{Zhang2004}
\bibinfo{author}{\bibfnamefont{S.}~\bibnamefont{Zhang}} \bibnamefont{and}
  \bibinfo{author}{\bibfnamefont{Z.}~\bibnamefont{Li}}, \bibinfo{journal}{Phys.
  Rev. Lett.} \textbf{\bibinfo{volume}{93}}, \bibinfo{pages}{127204}
  (\bibinfo{year}{2004}).

\bibitem[{\citenamefont{Boulle et~al.}(2008)\citenamefont{Boulle, Kimling,
  Warnicke, Kl\"aui, R\"udiger, Malinowski, Swagten, Koopmans, Ulysse, and
  Faini}}]{Boulle2008}
\bibinfo{author}{\bibfnamefont{O.}~\bibnamefont{Boulle}},
  \bibinfo{author}{\bibfnamefont{J.}~\bibnamefont{Kimling}},
  \bibinfo{author}{\bibfnamefont{P.}~\bibnamefont{Warnicke}},
  \bibinfo{author}{\bibfnamefont{M.}~\bibnamefont{Kl\"aui}},
  \bibinfo{author}{\bibfnamefont{U.}~\bibnamefont{R\"udiger}},
  \bibinfo{author}{\bibfnamefont{G.}~\bibnamefont{Malinowski}},
  \bibinfo{author}{\bibfnamefont{H.~J.~M.} \bibnamefont{Swagten}},
  \bibinfo{author}{\bibfnamefont{B.}~\bibnamefont{Koopmans}},
  \bibinfo{author}{\bibfnamefont{C.}~\bibnamefont{Ulysse}}, \bibnamefont{and}
  \bibinfo{author}{\bibfnamefont{G.}~\bibnamefont{Faini}},
  \bibinfo{journal}{Phys. Rev. Lett.} \textbf{\bibinfo{volume}{101}},
  \bibinfo{pages}{216601} (\bibinfo{year}{2008}).

\bibitem[{\citenamefont{Garate et~al.}(2009)\citenamefont{Garate, Gilmore,
  Stiles, and MacDonald}}]{Garate2009}
\bibinfo{author}{\bibfnamefont{I.}~\bibnamefont{Garate}},
  \bibinfo{author}{\bibfnamefont{K.}~\bibnamefont{Gilmore}},
  \bibinfo{author}{\bibfnamefont{M.~D.} \bibnamefont{Stiles}},
  \bibnamefont{and} \bibinfo{author}{\bibfnamefont{A.~H.}
  \bibnamefont{MacDonald}}, \bibinfo{journal}{Phys. Rev. B}
  \textbf{\bibinfo{volume}{79}}, \bibinfo{pages}{104416}
  (\bibinfo{year}{2009}).

\end{thebibliography}
\end{document}